# The adiabatic static linear response function in nonextensive statistical mechanics

Guo Ran and Du Jiulin

*Department of Physics, School of Science, Tianjin University, Tianjin 300072, China*

**Abstract** We analyze the difference between the three generations of the energy constraint in nonextensive statistical mechanics. Using the third generation of the energy constraint, we revise the isothermal static linear response function and then derive the adiabatic static linear response function under the adiabatic condition. We present the relationship between the isothermal and adiabatic linear response functions.

## 1. Introduction

The power-law distributions widely exist in the complex systems in nature and society, and have attracted considerable attention in recent years. In many different fields of scientific research, the power-law distributions have been frequently observed and studied, such as astronomy and astrophysics [1-6], plasma physics and space science [7-16], chemistry and life science [17-24]. Because the description of power-law distributions is beyond the scope of the traditional theory governed by Boltzmann-Gibbs (BG) statistics, a new statistical theory, nonextensive statistical mechanics (NSM) which is as a generalization of BG statistics [25], has been developed for the study of power-law distributions. Correspondingly, the stochastic dynamical theory for the power-law distributions has also been in progress [26-28]. We know that in certain circumstances, the power-law distributions can represent the nonequilibrium stationary-state of a complex system.

In nonequilibrium statistical mechanics, the linear response theory of the stationary state has been a powerful tool for the study of the response of the system affected under a weak external force [29]. The study of the linear response theory in NSM therefore became an interesting question. Chame *et al* studied the static linear response in NSM under an isothermal condition [30], and then Rajagopal worked the isothermal dynamical linear response theory [31]. In these works they employed the first or second generation of the energy constraint in NSM which may be not suitable for today's view. What is more, they both worked only under the isothermal condition, and thus the linear response theory in NSM under the adiabatic condition is still unknown. The purpose of this work is to revisit the isothermal linear response theory using the third generation of the energy constraint in NSM and then to study the linear response theory under the adiabatic condition.

The paper is organized as follows. In section 2, we will present a brief review on the three generations of the energy constraint in NSM and the differences between them. In section 3, the isothermal linear response function will be revised under the third generation of the energy constraint. In section 4, we will work out the linear response function under the adiabatic condition and then discuss the relation between the response functions under the two conditions. Finally in section 5, we give the conclusion.

## 2. A brief review of the three generations of energy constraints in NSM

In the quantum theory, Tsallis entropy can be generally defined [25] as

$$S_q = k_B \frac{1 - Tr\hat{\rho}^q}{q-1}, \tag{1}$$

where $k_B$ is Boltzmann constant, $\hat{\rho}$ is density matrix, and $q \neq 1$ is the nonextensive parameter. Maximizing Tsallis entropy subject to the normalization condition and the choice of the energy constraint, one could obtain the density matrix and define the $q$-expectation value for a physical quantity. With the development of NSM, three generations of the energy constraint were studied and appointed [25]. The first one is traditional,

$$U_q^{(1)} = Tr(\hat{H}\hat{\rho}), \tag{2}$$

the second one is $q$-dependent,

$$U_q^{(2)} = Tr(\hat{H}\hat{\rho}^q), \tag{3}$$

and the third one is $q$-normalized,

$$U_q^{(3)} = \frac{Tr(\hat{H}\hat{\rho}^q)}{Tr\hat{\rho}^q}, \tag{4}$$

where $\hat{H}$ is Hamiltonian of the system, and the superscripts (1), (2) and (3) distinguish the three energy constraints. With the three generations of the energy constraints, one can write the density matrices as a unified form [32],

$$\hat{\rho} = \frac{\left[1-(1-q^*)\beta^*\hat{H}\right]^{\frac{1}{1-q^*}}}{Tr\left[1-(1-q^*)\beta^*\hat{H}\right]^{\frac{1}{1-q^*}}}. \tag{5}$$

And they are equivalent to each other using the following parameter transformations, for the first constraint Eq.(2),

$$q^* = 2 - q^{(1)}, \tag{6}$$

$$\beta^* = \frac{\beta^{(1)}}{q^{(1)}Tr\hat{\rho}^{q^{(1)}} + (q^{(1)}-1)\beta^{(1)}U_q^{(1)}}, \tag{7}$$

for the second constraint Eq.(3),

$$q^* = q^{(2)}, \tag{8}$$

$$\beta^* = \beta^{(2)}, \tag{9}$$

and for the third constraint Eq.(4),

$$q^* = q^{(3)}, \tag{10}$$

$$\beta^* = \frac{\beta^{(3)}}{Tr\hat{\rho}^{q^{(3)}} + (1-q^{(3)})\beta^{(3)}U_q^{(3)}}, \tag{11}$$

where $\beta^{(i)}$ with $i=1, 2, 3$ corresponds to the Lagrange multipliers for the energy constraints of Eq.(2), Eq.(3) and Eq.(4) respectively.

On the other hand, these different energy constraints lead to different definitions of $q$-expectation value for an arbitrary physical quantity $\hat{O}$, such as

$$\langle \hat{O} \rangle_q^{(1)} = Tr(\hat{O}\hat{\rho}) = \frac{Tr\left\{\hat{O}\left[1-(1-q^*)\beta^*\hat{H}\right]^{\frac{1}{1-q^*}}\right\}}{Tr\left[1-(1-q^*)\beta^*\hat{H}\right]^{\frac{1}{1-q^*}}}, \quad (12)$$

$$\langle \hat{O} \rangle_q^{(2)} = Tr(\hat{O}\hat{\rho}^{q^{(2)}}) = \frac{Tr\left\{\hat{O}\left[1-(1-q^*)\beta^*\hat{H}\right]^{\frac{q^*}{1-q^*}}\right\}}{\left\{Tr\left[1-(1-q^*)\beta^*\hat{H}\right]^{\frac{1}{1-q^*}}\right\}^{q^*}}, \quad (13)$$

$$\langle \hat{O} \rangle_q^{(3)} = \frac{Tr(\hat{O}\hat{\rho}^{q^{(3)}})}{Tr\hat{\rho}^{q^{(3)}}} = \frac{Tr\left\{\hat{O}\left[1-(1-q^*)\beta^*\hat{H}\right]^{\frac{q^*}{1-q^*}}\right\}}{Tr\left[1-(1-q^*)\beta^*\hat{H}\right]^{\frac{q^*}{1-q^*}}}, \quad (14)$$

where Eqs.(5)-(11) have been used. Therefore the $q$-expectation values, Eqs.(12), (13) and (14), are not equivalent to each other.

It is worth noticing that Eq.(12) can turn into Eq.(14) with the transformation $q^* \to 2-1/q^*$ and $\beta^* \to q^*\beta^*$, but Eq.(13) can never turn into Eq.(14) with any transformation. So far, one has not found any unique transformation that can make both the three density matrices and the three $q$-expectation values to transform each other. Using the transformation Eqs.(6)-(11) one can not write the three $q$-expectation values, Eqs.(12)-(14), as one unified form. In the traditional statistical mechanics, there exists only one definition for the expectation value and thus the density matrix contains all information of the system. In NSM, because there are at least three definitions for the $q$-expectation value, the use of different definitions can lead to different results. For example, the relative energy fluctuation for a classical ideal gas in the canonical ensemble of NSM is, for the second constraint [33],

$$\frac{\sqrt{\langle E-U_q^{(2)} \rangle_q^{(2)}}}{U_q^{(2)}} = \sqrt{\frac{2}{3q^{(2)}N} + \frac{1-q^{(2)}}{q^{(2)}}}, \quad (15)$$

but for the third constraint [34],

$$\frac{\sqrt{\left\langle E-U_q^{(3)}\right\rangle_q^{(3)}}}{U_q^{(3)}}=\frac{2}{3N}\sqrt{\frac{2}{3-q^{(3)}}\frac{3N}{(1-q^{(3)})3N+2}}. \quad (16)$$

If the transformation equations (8) and (10) are applied to Eq.(15) and Eq.(16), i.e. $q^{(2)}=q^{(3)}$, clearly we find that these two expressions are entirely different. In other words, the three generations of the energy constraint, Eqs.(2), (3) and (4), are generally not equivalent. When we make an evaluation of the expectation value for a physical quantity in NSM, we have to choose a suitable energy constraint to define the $q$-expectation value.

### 3. The isothermal static linear response function revisited

Generally speaking, an $n$-particle system can be described by the Hamiltonian $H(\mathbf{q},\mathbf{p})$ where $(\mathbf{q},\mathbf{p})=(\{q_i\}_{i=1,2,...,n}, \{p_i\}_{i=1,2,...,n})$. If the system is acted by a small external force $\delta x$, then the Hamiltonian becomes

$$\hat{H}_\delta(\mathbf{q},\mathbf{p},\delta x)=\hat{H}(\mathbf{q},\mathbf{p})-\hat{A}(\mathbf{q},\mathbf{p})\delta x, \quad (17)$$

where $\hat{A}(\mathbf{q},\mathbf{p})$ is the corresponding displacement and $\delta x$ is time-independent for the static response. The linear response indicates that the density matrix and any of observable physical quantity $\hat{O}$ of the system will respond the perturbation with the following equation [29],

$$\left\langle \hat{O}(\delta x)\right\rangle-\left\langle \hat{O}(0)\right\rangle=\chi_{OA}\delta x, \quad (18)$$

where $\chi_{OA}$ is called static linear response function or static susceptibility.

We now study the isothermal static response function under the third generation of the energy constraint in NSM. The isothermal static response function was studied in [30] and [31], both utilizing the second generation of the energy constraint Eq.(3). According to today's point of view, the third generation of the energy constraint Eq.(4) has more advantage than the second one [25]. It is therefore necessary to revise the isothermal static response function with Eq.(4).

If the system is governed by NSM, substituting Eq. (17) into Eq. (5) and Eq.(14), one can derive the density matrix,

$$\hat{\rho}(\delta x) = \frac{\left[1-(1-q^*)\beta^*\left(\hat{H}-\hat{A}\delta x\right)\right]^{\frac{1}{1-q^*}}}{Tr\left[1-(1-q^*)\beta^*\left(\hat{H}-\hat{A}\delta x\right)\right]^{\frac{1}{1-q^*}}}, \quad (19)$$

and the $q$-expectation value for an arbitrary physical quantity $\hat{O}$

$$\langle\hat{O}(\delta x)\rangle_q^{(3)} = \frac{Tr\left[\hat{O}\hat{\rho}^{q^*}(\delta x)\right]}{Tr\hat{\rho}^{q^*}(\delta x)}$$

$$= \frac{Tr\left\{\hat{O}\left[1-(1-q^*)\beta^*\left(\hat{H}-\hat{A}\delta x\right)\right]^{\frac{q^*}{1-q^*}}\right\}}{Tr\left[1-(1-q^*)\beta^*\left(\hat{H}-\hat{A}\delta x\right)\right]^{\frac{q^*}{1-q^*}}}, \quad (20)$$

where $\beta^*$ and $q^*$ are given in Eqs.(10) and (11). In the isothermal case, the inverse temperature $\beta^*$ is not affected by the perturbation $\delta x$.

The $q$-expectation Eq.(20) can also be expressed using the escort density matrix $\hat{P}$ [25] as

$$\langle\hat{O}(\delta x)\rangle_q^{(3)} = Tr\left[\hat{O}\hat{P}(\delta x)\right], \quad (21)$$

with the definition,

$$\hat{P}(\delta x) = \frac{\hat{\rho}^{q^*}(\delta x)}{Tr\hat{\rho}^{q^*}(\delta x)}. \quad (22)$$

In this definition, $q^* \equiv q^{(3)}$, for convenience to write hereinafter, we will omit the superscript of $q$. Following Kubo's line [29] concisely, we use the identical equation [31],

$$\left[1-(1-q)\beta^*\left(\hat{H}-\hat{A}\delta x\right)\right]^{\frac{q}{1-q}}$$

$$= \left[1-(1-q)\beta^*\hat{H}\right]^{\frac{q}{1-q}}\left\{1+q\delta x\int_0^{\beta^*}d\lambda\left[1-(1-q)\lambda\left(\hat{H}-\hat{A}\delta x\right)\right]^{\frac{q}{1-q}-1}\hat{A}\left[1-(1-q)\lambda\hat{H}\right]^{\frac{q}{1-q}-1}\right\}. \quad (23)$$

And then substituting Eq.(23) into Eq.(19), expanding it for $\delta x$ and retaining the first-order approximation, we can write the escort density matrix as

$$\frac{\hat{\rho}^q(\delta x)}{Tr[\hat{\rho}^q(\delta x)]} = \frac{\hat{\rho}^q(0)}{Tr[\hat{\rho}^q(0)]}\left(1+q\beta^*\delta x\Delta\hat{A}'\right), \quad (24)$$

where

$$\Delta\hat{A}' = \hat{A}' - \langle\hat{A}'\rangle_q^{(3)}, \quad (25)$$

and

$$\hat{A}' = \beta^{*-1}\int_0^{\beta^*} d\lambda\left[1-(1-q)\lambda\hat{H}\right]^{-\frac{q}{1-q}-1}\hat{A}\left[1-(1-q)\lambda\hat{H}\right]^{\frac{q}{1-q}-1}. \quad (26)$$

In the classical limit: $\hbar \to 0$, Eq.(26) is reduced to

$$A' = \frac{A}{1-(1-q)\beta^* H}. \quad (27)$$

On the right-hand side of Eq.(27), the mathematical structure of the denominator is from the power-law distribution. In the limit: $q \to 1$, Eq.(26) becomes

$$\hat{A}' = \beta^{*-1}\int_0^{\beta^*} d\lambda \exp(\lambda\hat{H})\hat{A}\exp(-\lambda\hat{H}), \quad (28)$$

and Eq.(27) is $A' = A$, both recovering the traditional forms in [29].

Multiplying by $\hat{O}$ on both sides of Eq.(24) and then taking the trace of matrix, we find that

$$\langle\hat{O}(\delta x)\rangle_q^{(3)} - \langle\hat{O}(0)\rangle_q^{(3)} = q\beta^*\langle\Delta\hat{A}'\Delta\hat{O}\rangle_q^{(3)}\delta x, \quad (29)$$

where $\Delta\hat{O}$ is defined the same as $\Delta\hat{A}'$ in Eq.(25), i.e.

$$\Delta\hat{O} = \hat{O} - \langle\hat{O}\rangle_q^{(3)}.$$

According to the definition of the linear response function in the equation,

$$\langle\hat{O}(\delta x)\rangle_q^{(3)} - \langle\hat{O}(0)\rangle_q^{(3)} = \chi_{OA}\delta x, \quad (30)$$

we find that the isothermal static linear response function is

$$\chi_{OA}^T = q\beta^*\langle\Delta\hat{A}'\Delta\hat{O}\rangle_q^{(3)}, \quad (31)$$

or equivalently, it can be written as

$$\chi_{OA}^T = q\beta^*\left(\langle\hat{A}'\hat{O}\rangle_q^{(3)} - \langle\hat{A}'\rangle_q^{(3)}\langle\hat{O}\rangle_q^{(3)}\right). \quad (32)$$

The results based on the second constraint in [30] and [31] can be re-arranged by

$$\chi_{OA}^{T} = q^{*}\beta^{*}\left(\left\langle \hat{A}'\hat{O}\right\rangle_{q^{*}}^{(3)} - \left\langle \hat{A}'\right\rangle_{2-q^{*}}^{(1)}\left\langle \hat{O}\right\rangle_{q^{*}}^{(3)}\right). \tag{33}$$

Now with $\beta^{*} = \beta^{(2)}$ and $q^{*} = q^{(2)}$, clearly we find that by using the transformation Eqs.(8)-(10), Eq.(32) and (33) cannot transform each other. This conclusion is consistent with that in Section 2.

## 4. The adiabatic static linear response function in NSM

In the adiabatic case, we will still follow the line in Ref.[31] to derive the linear response function in NSM. Unlike the isothermal case, the inverse temperature will have the small change $\delta\beta^{*}$ when the system is affected by an external force $\delta x$. Thus in the adiabatic case, the $q$-expectation value of the physical quantity turns into

$$\left\langle \hat{O}(\delta x, \delta\beta^{*})\right\rangle_{q}^{(3)} = \frac{Tr\left\{\hat{O}\left[1-(1-q)(\beta^{*}+\delta\beta^{*})(\hat{H}-\hat{A}\delta x)\right]^{\frac{q}{1-q}}\right\}}{Tr\left[1-(1-q)(\beta^{*}+\delta\beta^{*})(\hat{H}-\hat{A}\delta x)\right]^{\frac{q}{1-q}}}, \tag{34}$$

and the escort density matrix is, with the first-order approximation,

$$\frac{\hat{\rho}^{q}(\delta x, \delta\beta^{*})}{Tr\left[\hat{\rho}^{q}(\delta x, \delta\beta^{*})\right]} = \frac{\hat{\rho}^{q}(0,0)}{Tr\left[\hat{\rho}^{q}(0,0)\right]}\left(1-q\delta\beta^{*}\Delta\hat{H}' + q\delta x\Delta\hat{A}'\right), \tag{35}$$

where

$$\Delta\hat{H}' = \hat{H}' - \left\langle \hat{H}'\right\rangle_{q}^{(3)}, \tag{36}$$

and the integral in $\hat{H}'$ can be calculated,

$$\hat{H}' = \beta^{*-1}\int_{0}^{\beta^{*}} d\lambda\left[1-(1-q)\lambda\hat{H}\right]^{\frac{q}{1-q}-1}\hat{H}\left[1-(1-q)\lambda\hat{H}\right]^{\frac{q}{1-q}-1}$$
$$= \frac{\hat{H}}{1-(1-q)\beta^{*}\hat{H}}. \tag{37}$$

In the adiabatic case, the change of the Hamiltonian is totally from the work made by the external force, and therefore we have the adiabatic condition [29],

$$\delta\left\langle \hat{H}\right\rangle_{q}^{(3)} + \left\langle \hat{A}\right\rangle_{q}^{(3)}\delta x = 0, \tag{38}$$

where $\hat{H}$ still denotes $\hat{H}(\mathbf{p},\mathbf{q})$. The change of the Hamiltonian is calculated by

$$\delta \langle \hat{H} \rangle_q^{(3)} = \frac{Tr\left[\left(\hat{H} - \hat{A}\delta x\right)\rho^q\left(\delta x, \delta \beta^*\right)\right]}{Tr\left[\rho^q\left(\delta x, \delta \beta^*\right)\right]} - \frac{Tr\left[\hat{H}\rho^q(0,0)\right]}{Tr\left[\rho^q(0,0)\right]}. \tag{39}$$

Also retaining the first-order approximation for $\delta x$ and $\delta \beta^*$ in Eq.(39), it becomes that

$$\delta \langle \hat{H} \rangle_q^{(3)} = -\langle \hat{A} \rangle_q^{(3)} \delta x - q \langle \Delta \hat{H}' \Delta \hat{H} \rangle_q^{(3)} \delta \beta^* + \beta^* q \langle \Delta \hat{H}' \Delta \hat{A} \rangle_q^{(3)} \delta x. \tag{40}$$

Comparing Eq.(38) with Eq.(40), we can find the relation between $\delta x$ and $\delta \beta^*$,

$$\frac{\delta \beta^*}{\beta^*} = \frac{\langle \Delta \hat{H}' \Delta \hat{A} \rangle_q^{(3)}}{\langle \Delta \hat{H}' \Delta \hat{H} \rangle_q^{(3)}} \delta x. \tag{41}$$

Substituting Eq.(41) into Eq.(35), we obtain the adiabatic linear response function,

$$\chi_{OA}^{ad} = \beta^* \left( q \langle \Delta \hat{A}' \Delta \hat{O} \rangle_q^{(3)} - \frac{\langle \Delta \hat{H}' \Delta \hat{A} \rangle_q^{(3)} \langle \Delta \hat{H}' \Delta \hat{O} \rangle_q^{(3)}}{\langle \Delta \hat{H}' \Delta \hat{H} \rangle_q^{(3)}} \right), \tag{42}$$

where the first term in Eq.(42) is exactly the isothermal response function $\chi_{OA}^T$ in Eq.(31), and the second term is naturally the effect of temperature fluctuation caused by the small external force. In the limit $q \to 1$ and $\beta^* = \beta$, Eq.(42) recovers to the traditional form in [29],

$$\chi_{OA}^{ad} = \beta \left( \langle \Delta \hat{A}; \Delta \hat{B} \rangle - \frac{\langle \Delta \hat{H} \Delta \hat{A} \rangle \langle \Delta \hat{H} \Delta \hat{B} \rangle}{\langle \Delta \hat{H} \Delta \hat{H} \rangle} \right), \tag{43}$$

where the canonical correlation function is

$$\langle \Delta \hat{A}; \Delta \hat{B} \rangle = \beta^{-1} \left\langle \int_0^\beta d\lambda \exp(\lambda \hat{H}) \Delta \hat{A} \exp(-\lambda \hat{H}) \Delta \hat{B} \right\rangle. \tag{44}$$

Noting that due to

$$\frac{\partial}{\partial \beta^*} \langle \hat{O} \rangle_q^{(3)} = \frac{\partial}{\partial \beta^*} \frac{Tr\left\{\hat{O}\left[1-(1-q)\beta^*\hat{H}\right]^{\frac{q}{1-q}}\right\}}{Tr\left[1-(1-q)\beta^*\hat{H}\right]^{\frac{q}{1-q}}} = -\langle \Delta \hat{H}' \Delta \hat{O} \rangle_q^{(3)}, \tag{45}$$

Eq.(42) can be transformed into a thermodynamic equation as that in the conventional statistics, i.e.

$$\chi_{OA}^{ad} = \chi_{OA}^{T} - \beta^{*} \frac{\frac{\partial}{\partial \beta^{*}} \langle \hat{A} \rangle_{q}^{(3)} \frac{\partial}{\partial \beta^{*}} \langle \hat{O} \rangle_{q}^{(3)}}{\frac{\partial}{\partial \beta^{*}} \langle \hat{H} \rangle_{q}^{(3)}}. \tag{46}$$

If we consider the real physical temperature to be $T = 1/k\beta^{*}$, and the $q$-expectation value to be a physical quantity that can be measured in experiments, the relation between the isothermal and adiabatic response functions, $\chi_{OA}^{T}$ and $\chi_{OA}^{ad}$, is exactly the same as that in the traditional statistics. However, what is the physical temperature in NSM? It is still under discussion and this topic is beyond the present paper. We can present an example explained for the physical temperature in [35],

$$T_{phys} = \frac{Tr\rho^{q}}{k\beta^{(3)}} = \frac{1}{k\beta^{*}} - (1-q)\frac{U_{q}^{(3)}}{k}. \tag{47}$$

If this is the temperature in an real physical system, the relation (46) turns into

$$\chi_{OA}^{ad} = \chi_{OA}^{T} - \frac{T_{phys} + (1-q)\frac{U_{q}^{(3)}}{k}}{C_{q}} \frac{\partial}{\partial T_{phys}} \langle \hat{A} \rangle_{q}^{(3)} \frac{\partial}{\partial T_{phys}} \langle \hat{O} \rangle_{q}^{(3)}, \tag{48}$$

where $C_{q}$ is the heat capacity,

$$C_{q} = \frac{\partial}{\partial T_{phys}} \langle \hat{H} \rangle_{q}^{(3)}. \tag{49}$$

It is impressed that the different definitions for the physical temperature will not change both forms of the isothermal and adiabatic response functions, Eq.(31) and Eq.(42), because there is always a functional relation between the temperature and $\beta^{*}$ like Eq.(47). In the isothermal case, $\beta^{*}$ does not change, while in the adiabatic case, $\beta^{*}$ has a small change as the temperature changes.

## 5. Conclusion

In conclusion, first we have shown that the three generations of the energy constraint equations, (2)~(4), in NSM are not equivalent, which means that we will obtain the different results if we use the different energy constraint. The reason is that the different energy constraint gives the different definition for the *q*-expectation value of an arbitrary observable.

Second, employing the third generation of the energy constraint we have revised the static linear response function in NSM under the isothermal condition. The new static linear response function, Eq.(32), is thus different from those in [30] and [31].

Third, employing the third generation of the energy constraint we have obtained the static linear response function, Eq.(42), in NSM under the adiabatic condition. And then we have found the relation, Eq.(46), between the isothermal and adiabatic response functions.

Finally, by an example we have shown that the different definition for the physical temperature does not change the form of both the isothermal and adiabatic response functions.

**Acknowledgements**

This work is supported by the National Natural Science Foundation of China under Grant No. 11175128 and by the Higher School Specialized Research Fund for Doctoral Program under Grant No. 20110032110058.